\documentclass[]{elsarticle}
\usepackage{amsmath,amssymb,amsfonts}
\usepackage{algorithmic}
\usepackage{graphicx}
\usepackage{textcomp}
\usepackage{xcolor}
\usepackage{balance}
\usepackage{multirow}

\def\BibTeX{{\rm B\kern-.05em{\sc i\kern-.025em b}\kern-.08em
    T\kern-.1667em\lower.7ex\hbox{E}\kern-.125emX}}
\begin{document}

\title{Rewrite it in Rust: A Computational Physics Case Study}

\author[1]{Willow Veytsman}
\ead{wveytsma@u.rochester.edu}
\author[1]{Shuang Zhai}
\ead{szhai2@cs.rochester.edu}
\author[1]{Chen Ding}
\ead{cding@cs.rochester.edu}
\author[1]{Adam B. Sefkow}
\ead{adam.sefkow@rochester.edu}
\affiliation[1]{
    organization={University of Rochester},
    city={Rochester},
    country={USA}
}

\begin{abstract}
Surveys of computational science show that many scientists use languages like C and C++ in order to write code for scientific computing, especially in scenarios where performance is a key factor. In this paper, we seek to evaluate the use of Rust in such a scenario, through implementations of a physics simulation in both C++ and Rust. We also create a parallel version of our Rust code, in order to further explore performance as well as parallel code complexity. Measuring performance as program runtime, we find that Rust can offer better performance than C++, with some test cases showing as much as a 5.6$\times$ performance increase, and that parallel code in Rust can further improve performance while being easy to write safely. Finally, we provide some preliminary profiling to better understand the difference between the way our implementations perform.
\end{abstract}


\maketitle
\pagestyle{plain}

\section{Introduction}

As the field of computational science continues to grow, demand increases for the use of high-performance scientific computation, which enables faster modeling of more complex interactions and analysis of larger datasets. A survey of computational scientists across a wide range of disciplines conducted by Princeton University found that close to 1/4 of the scientists use programming languages like C and C++ in order to write performance-intensive computation code~\cite{Prabhu+:SC11}. Further surveys of computational science show similar results: an international survey of research software engineers in academia done in 2018 found that about 1/3 of scientists use C and/or C++ in their code~\cite{Philippe+:2019}.

Unlike MATLAB and Python, C and C++ are compiled and rely on manual memory management rather than garbage collection, offering the greater performance needed in certain applications. They are established programming languages, having been around for several decades, with numerous libraries and tools written for various computational science applications.
Rust is a relative newcomer in the programming language space. It offers similar performance to languages such as C and C++, but takes a variety of new approaches to solve problems such as memory safety and parallelism, making it a great choice for computational scientists.

This paper looks at the benefits of Rust through the re-implementation of a simulation of laser-plasma interaction from C++ to Rust. Specifically, the Cross Beam Energy Transfer (CBET) effect happens when high-intensity laser beams intersect in a plasma. The measurable outcome of the interaction is a change in the beams' intensities, i.e. how much energy is present in each participating laser beam after the interaction is calculated. Simulating this interaction is a crucial step in building an accurate simulator of laser systems in order to aid research on topics such as inertial confinement nuclear fusion, high-energy-density physics, materials science, laboratory astrophysics, and fundamental science. 

\section{Background}
\label{sec:bg}
\subsection{The CBET Problem}
\label{sec:bg:cbet}
Researchers worldwide are studying fusion energy for its potential to be a clean and virtually unlimited source of electricity and industrial heat in the future.
For laser-driven fusion studies, one of the biggest challenges is to model various factors that affect the fusion efficiency, including the Cross Beam Energy Transfer (CBET) effect~\cite{Randall+:PoF81}.
CBET is an interaction that happens when multiple laser beams intersect in plasma. When they intersect, they can interact with plasma waves and a certain amount of energy can be transferred from one beam to another; the same concept generalizes from two beams to N beams intersecting.
While the use of experimental laser systems is in high demand, they are expensive to operate and slow, so building a fast and accurate simulator would make it easier for scientists to design experiments to run on laser systems, advancing the speed with which research can be done on them.

The CBET program represents the plasma-filled space that the laser beams pass through as a mesh, which discretizes the space into small regions called zones. Each zone of the mesh corresponds to a specific subsection of the space, and carries information about that space, such as its electron density. The presence of an electron density gradient in space causes the rays to refract (i.e. change direction) within the plasma. Laser beams are represented as a set of discrete rays, with each ray's trajectory represented as a list of crossings between that ray and the zones of the mesh. Each crossing stores the intensity of that ray at that crossing.

The following is a short summary of the simulator code:
\begin{enumerate}
    \item Create the mesh, beams, and rays.
    \item Trace the rays: for each ray, follow its trajectory and save the location of each crossing of the mesh. Set the intensity of each crossing to the initial intensity of the laser beam.
    \item Calculate the CBET interaction multiplier: each pair of crossing rays from different beams is allowed to exchange energy according to an equation governing the interaction \cite{Follett+:PRE18}.
    \item Update the intensities: modify the ray intensities for the downstream portions of the trajectories for the crossing rays. Each crossing results in one of the rays increasing in intensity and the other decreasing.
    \item Repeat the previous two steps until the change in intensity is smaller than a user-prescribed ``convergence" constant. Typically a relative change of $10^{-3}$ to $10^{-4}$ is sufficient.
    \item Calculate the average intensity value of all of the crossings at each spot in the grid, in order to determine the resulting field amplitude for each zone.
\end{enumerate}

There are multiple academic teams and research facilities interested in tackling this problem using numerical models.
They implement various software programs and each has its own pros and cons.
Some examples are Lilac~\cite{Delettrez+:PRA87} DRACO~\cite{Radha+:POP05}, Aster~\cite{Igumenshchev+:POP16}, and Hydra~\cite{Marinak+:POP96}.
Follett et al~\cite{Follett+:POP22} implemented the ray-based CBET model discussed in this paper first in MATLAB for expediency without consideration to performance.
It is able to consider multiple physics phenomena and produce accurate results, which are verified by comparing against other simulation tools.
However, due to performance limitations with the programming language and lack of software engineering practices during development, the MATLAB version is only capable of running small problems.
Also, it is difficult to scale the code to run on a server with hundreds of cores or even on accelerators. 
Therefore, it is important to study how different programming languages could improve performance, scalability, and correctness when building large scientific programs like CBET.

\subsection{Rust}
\label{sec:bg:rust}
Rust~\cite{KlabnikN:2017} is a programming language that takes some new approaches to certain problems in programming. Most important for this paper are its approaches to memory management and parallelism.

Memory management has traditionally been addressed through manual memory management, as in C or C++, or through garbage collection, such as in MATLAB and Python. Garbage collection has significant performance costs, whereas manual memory management can introduce a lot of potential errors when not done carefully~\cite{Scott:PLPv4}. Rust's approach is called the Ownership System. It prevents most of the errors manual memory management can create by enforcing a few rules on variables and references to them:

\begin{itemize}
    \item Each value (stored in memory) has exactly one owner variable. The memory is deallocated (dropped) when the owner goes out of scope. Ownership can be moved from one variable to another.
    \item Unlimited immutable references (also called borrows) can be made for any variable. However, as long as these immutable references are in scope, the memory they refer to cannot be mutated.
    \item One mutable reference (also called a mutable borrow) can be made for any variable. As long as this reference is in scope, the memory can only be read from or written to through that reference.
\end{itemize}

These are called ownership and borrow rules~\cite{KlabnikN:2017}.  Because these rules can be checked at compile-time, they offer solutions to memory problems that have no runtime cost. This makes Rust an attractive option for writing code where high-performance is needed, such as CBET. More language features that are useful to CBET are introduced in Sec~\ref{sec:impl:rust}.

\section{Implementing CBET}
\label{sec:impl}
The following subsections describe the implementations of the CBET code in MATLAB, C++, and Rust, noting the impacts of the programming language, as well as providing more details on how the rough outline described above was put into code.

\subsection{C++ and MATLAB Implementation}
\label{sec:impl:cpp}
The C++ version of CBET is implemented based on the MATLAB code written by Follett et al~\cite{Follett+:POP22,Follett+:PRE18}.
A laser beam to be launched into plasma is discretized by a set of initially parallel rays.
The computational modeling process involves two major steps of tracking rays across a mesh, which represents the plasma conditions (density, temperature, flow velocity, etc.).

The first major step is the ray tracing phase, wherein we must calculate the trajectory of each ray using coupled differential equations. There are several different forms the equations can take, but the form we chose and find to be most understandable is
\begin{equation}
\frac{d}{dt} \textbf{v} = -\frac{c^2}{2 n_{crit}} \nabla n_e
\end{equation}
\begin{equation}
\frac{d}{dt} \textbf{x} = \textbf{v}.
\end{equation} 
At each time step, the ray's position vector (\textbf{x}) is updated based on the velocity vector (\textbf{v}), whose calculation depends on the electron density gradient $\nabla n_e$ calculated at the current mesh location of the ray. The variable $c$ is the speed of light and $n_{crit}$ is the critical density of the plasma which depends on the wavelength of the laser light.
The dispersion relation that is enforced for transverse electromagnetic waves (i.e. laser light) is
\begin{equation}
\omega_L^2 = \omega_{p,e}^2 + c^2 k^2,
\end{equation}
where $\omega_L$ is the laser frequency, $\omega_{p,e}$ is the electron plasma frequency, and $k$ is the wavevector.
The wavevector, laser frequency, and velocity of the ray are related by the equation
\begin{equation}
\textbf{v} = c^2 \frac{\textbf{k}}{\omega_L}.
\end{equation}
The net result of this process is that each of the many rays representing each laser beam is marched across the mesh. Typically each ray travels in slightly different directions with slightly different headings and velocities. The entire path traced out is called the ray's trajectory, which typically bends (i.e. refracts) in the presence of the electron density gradient. Each ray trajectory calculation is independent of the others and so this phase should be embarassingly parallel. 
After every ray exits the mesh, we have a complete trajectory for each ray.
We can also derive other important values from the ray trajectories, such as \textbf{k} vectors when rays enter new zones.
These values will be used to calculate the CBET in the second step.
We also launch an ``auxiliary" ray (or child ray) with a slight offset compared to the corresponding main ray (or parent ray).
The child ray has the same intensity and initial \textbf{k} vector as its parent ray.
As both rays go through the plasma, the subtle difference in initial position will give them a slightly different velocity at each time step, and therefore the trajectories are not perfectly parallel.
The change in separation distance between main ray and auxiliary ray represent the change in the laser beam's intensity, $I_L$ [W/m$^{2}$]. A laser's intensity is related to its electric field strength $E$ [V/m] through the relation 
\begin{equation}
I_L = \frac{1}{2} \epsilon_0 c E^2
\end{equation}
where $\epsilon_0$ is the permittivity of free space. An equivalent way of characterizing the strength of the laser field is the normalized vector potential
\begin{equation}
a_0 = \frac{e E}{m_e c \omega_L} = \sqrt{\frac{2 I_L}{\epsilon_0 c^3}} \frac{e}{m_e \omega_L}
\end{equation}
where $e$ and $m_e$ are the charge and mass of the electron, respectively. As equation (6) illustrates, $a_0$, $E$, and $I_L$ are all related and interchangeable when discussing the electromagnetic field strength of the laser.

The second major step is the CBET calculation itself, which is responsible for evaluating the redirection of laser intensity or energy among the rays making up each laser beam.
Based on the trajectories obtained from step one, if two rays enter the same zone, they are considered to intersect, and therefore energy can be exchanged.
The overall CBET calculation step is implemented as an iterative procedure: for each ray, the code iterates over all possible intersections, calculates the CBET coupling multiplier for energy exchange, and propagates the result of the energy change to the remaining downstream portion of the trajectory.
The process is repeated until the solution converges, meaning the answer stops changing with additional iterations (i.e, the maximum amount of relative energy exchange within a single iteration is below a predefined, small threshold; typical values might be 1e-3 to 1e-4).


Although the C++ version is based on the MATLAB code, there are still important differences between the two implementations.
First, the design goals are different; the MATLAB code is designed with the purpose of simulating a large variety of problems and validating the results against existing tools.
Therefore, the implementation is more focused on feature completion and accuracy.
Meanwhile, the C++ code is optimized for efficiency.
It contains a subset of features from MATLAB that are essential to the ICF CBET modeling, but it is rewritten to run faster and more efficiently.
Second, the two programming languages offer different features in terms of memory management.
C++ has explicit memory allocation/deallocation interfaces, which makes efficient memory usage possible.
However, there is no counterpart in MATLAB, which makes certain optimizations impossible to implement.
For example, when tracing child rays and calculating the cross sectional area ratios, the MATLAB version needs to allocate arrays to store every child ray's trajectory and then compute the cross sectional area ratios.
The C++ version, on the contrary, only needs to allocate memory for the first ray trajectory, and immediately calculates the area ratios.
The same array is reused for subsequent child rays.
This technique is heavily adopted in the C++ code.
It substantially reduces the memory usage and improves the cache utilization.

\begin{figure}
    \centering
    \includegraphics[width=0.8\linewidth]{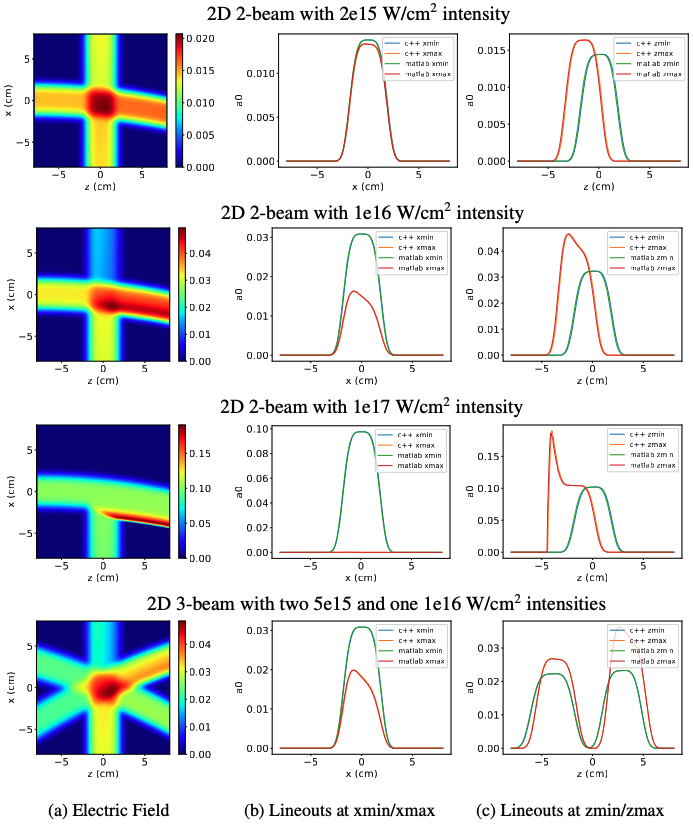}
    \caption{Test cases of the 2D 2-beam and 3-beam CBET problem using different initial laser intensities. All plots and lineouts are in the dimensionless units of the normalized vector potential $a_0$. The beam moving from bottom-to-top transfers its energy to the beam(s) moving from left-to-right; the amount increases as the intensity increases until the beam losing energy is fully depleted.}
    \label{fig:SAMM-basics}
\end{figure}

To make a fair comparison, the C++ version used in Sec~\ref{sec:perf} is modified to be more similar to the Rust implementation as described in Sec~\ref{sec:impl:rust}. However, the structure of the data is different in some cases.
Figure ~\ref{fig:SAMM-basics} shows the CBET results with different initial laser intensities and numbers of lasers.
All three implementations of the CBET code are capable of generating the same results.

\subsection{Rust Implementation}
\label{sec:impl:rust}

The Rust implementation of the CBET program was created after the C++ implementation, and was heavily based off the C++ code. The methods for performing the calculation are the same, but the design of the Rust implementation is somewhat different, taking advantage of Rust's modularity features. Rust treats each source code file as a module, which is given its own namespace when referenced by other parts of code. While C++ also has support for namespaces, Rust's system forces the developer to use them, ultimately producing code that is more readable and easier to maintain with little difficulty. By moving code into modules, we can also upload the CBET code to Rust's package repositories in the future for others to use.

Some important language features used in the Rust program are iterators and closures.  An \emph{iterator} is a control abstraction for performing a task on a sequence of items. What sets Rust apart from other programming languages that implement iterators is the extremely low overhead with which they are implemented in Rust~\cite[Chapter 13]{KlabnikN:2017}.
The task is written in an anonymous function, which is a first-class value that can be created dynamically based on variable values in a running function, returned from the function, or passed as a parameter when calling another function~\cite{Scott:PLPv4}.  In Rust, anonymous functions are implemented as closures.

Once a collection is turned into an iterator, it can be adapted with methods such as ``map'', or consumed with methods such as ``fold'' and ``reduce''. Iterator methods are lazy, meaning that they do not do anything until they are consumed. In addition to lazy evaluation, iterators also remove the need for loop control code.
Later we will evaluate the cost of using iterators and closures in CBET.  


In Rust, the ownership and borrowing system is also enforced on closures, and functions can specify whether or not the closure can borrow references from their environment, mutably or immutably, in their type signature~\cite[Chapter 13]{KlabnikN:2017}. This feature, in addition to contributing to memory safety, also makes parallelism easier, as functions which execute in parallel can specify that they can't borrow mutable references from their environment (which could cause data races), and code that does so will not compile. This is in contrast with C and C++, which will compile code that contains unsafe memory use, often requiring extended debugging to fix errors or assure a lack of them.

\begin{figure}[h!]
    \centering
    \includegraphics[width=1\linewidth]{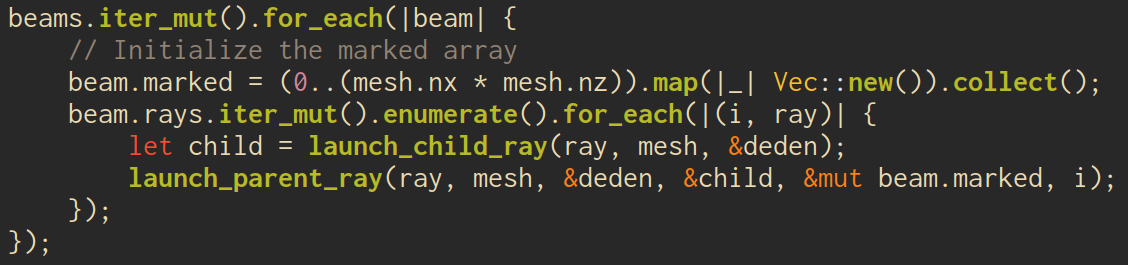}
    \caption{Rust Example: Iterators and Closures}
    \label{fig:seq_iter_ex}
\end{figure}

An example of the use of iterators and closures in the CBET program can be found in Fig~\ref{fig:seq_iter_ex}, which shows an excerpt from the ray tracing part of the code. In this example, the ``beams'' variable refers to a vector of beam structs. The ``iter\_mut'' method on the vector transforms it into an iterator, where each reference in the iterator is mutable. Then, the ``for\_each'' method runs the closure it is provided on each element of the iterator. The third line initializes the ``marked" array, which contains a vector for each zone in the mesh. It does so by creating a range, which is a type of iterator, and then using the ``map'' adapter to change each element in the range into an empty vector. Finally, the ``collect'' iterator method turns the iterator into a collection, here a vector. The following lines iterate over the rays of the beam in a similar way, with the ``enumerate'' adapter turning each element of the iterator into a tuple containing the index of the element and that element.

In order to explore the use of parallel code in Rust, a version of the Rust implementation was made that takes advantage of CPU-based multi-threading. The multi-threading is implemented using the Rayon library~\cite{Rayon}. Rayon builds on top of the Rust standard library's threading features to make parallel code easier to write and faster to execute. It creates a thread pool, and uses a work-stealing algorithm to distribute work evenly. While this creates a little bit of runtime overhead, as we show, the performance benefits of using Rayon in our CBET implementation are still significant.

Using the library can be as simple as replacing the use of methods which turn collections into iterators with Rayon's implementations, which turn collections into parallel iterators. Then, the methods implemented on the iterators, such as ``map'' and ``for\_each'', are run in parallel. The ease of this conversion is aided by the aforementioned rules around the closures that the iterator methods take. Rayon can specify in its function headers that the closures it takes must be thread-safe, and then the Rust compiler will refuse to compile code that isn't thread safe, drastically reducing the potential for thread-related errors such as data races. 

In order to guarantee memory safety at compile time, Rust's ownership and borrowing rules prevent some types of valid memory uses. For example, multiple threads may need to write to the same section of memory, but in thread-safe Rust code, mutable references may not be used in threads. In order to remedy this, Rust introduces so-called smart pointers, which enforce the borrow-checking rules at run-time instead of compile-time. One such smart pointer that is used in the parallel CBET implementation is the Mutex, which allows only one thread to read from or write to the memory it owns at any given time. Mutexes can be passed as immutable references to the closures that execute in parallel, and when they are locked, they give the thread access to a mutable reference to the memory inside.

\begin{figure}[h!]
    \centering
    \includegraphics[width=1\linewidth]{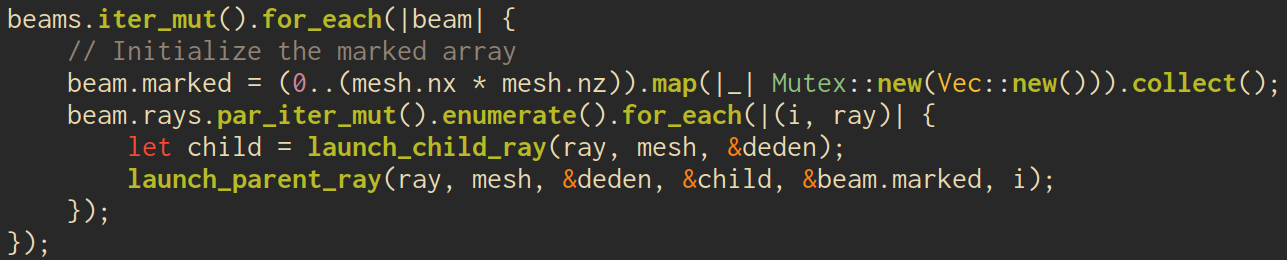}
    \caption{Rust Example: Parallel Iterators}
    \label{fig:par_iter_ex}
\end{figure}

An example of the parallel CBET code can be seen in Fig~\ref{fig:par_iter_ex}, which is the modified version of the earlier example in Fig~\ref{fig:seq_iter_ex}. Here, the iterator over the rays of each beam is modified to use Rayon's ``par\_iter\_mut'' method. Then, because the closure in the sequential code borrowed a reference to the marked array mutably, and that is not thread-safe, the borrow in the parallel code is not mutable, and instead Mutexes are used for memory that needs to be written to from multiple threads, a change that is reflected in the third line of the example.

In order to keep the comparison as direct as possible, no optimizations were made to the sequential Rust code besides using Rayon's parallel iterators and making the necessary modifications (such as adding Mutexes and modifying code to prevent writing to the same memory in multiple threads). The subroutines that were parallelized include ray tracing and the step of the CBET process that calculates the interaction multiplier. Making a version of the Rust implementation parallel involved only adding or modifying 67 source lines of code.

\section{Performance Results}
\label{sec:perf}

\subsection{Machine Specifications}
\label{sec:result:machine}

We ran the different implementations on three machines. They have been sorted in chronological order by processor, with the newest processors at the end. The Rust compiler version is 1.79.0 and the Clang compiler version is 14.0.6 on all machines. The first machine, referred to as Xeon, has the following specifications:
\begin{itemize}
    \item Intel Xeon W-2295 CPU
    \item Base frequency of 3.00 GHz, with boost up to 4.80 GHz
    \item 18 cores, 36 threads
    \item 24.75 MB L3 Cache
    \item 256 GB of memory
    \item GCC 11.4.0
\end{itemize}
The second machine, referred to as Threadripper, has the following specifications:
\begin{itemize}
    \item AMD Ryzen Threadripper PRO 5975WX CPU
    \item Base frequency of 3.6 GHz, with boost up to 4.5 GHz
    \item 32 cores, 64 threads
    \item 128 MB L3 Cache
    \item 128 GB of memory
    \item GCC 12.3.0
\end{itemize}
The third machine, referred to as EPYC, has the following specifications:
\begin{itemize}
    \item AMD EPYC 9124 CPU
    \item Base frequency of 3.00 GHz, with boost up to 3.7 GHz
    \item 16 cores, 32 threads
    \item 64 MB L3 Cache
    \item 192 GB of memory
    \item GCC 11.4.0
\end{itemize}

\subsection{Sequential Comparison}

We tested the MATLAB, C++ and Rust implementations of CBET, simulating the interaction of two high-intensity laser beams in a two-dimensional mesh. The C++ and Rust implementations were compiled with -O3 optimizations, and the Rust implementation was run with one thread, to make a fair sequential comparison. The C++ code was also compiled to target the CPUs it was being compiled on, using the -march=native compiler option. The benefit of using this option varied depending on the compiler and machine used, with performance improvements of up to 26\% being found. Rust has a similar compiler option, -Ctarget-cpu=native, but the use of that option did not significantly impact the performance of the Rust programs.

\begin{figure}
    \centering
    \includegraphics[width=0.75\linewidth]{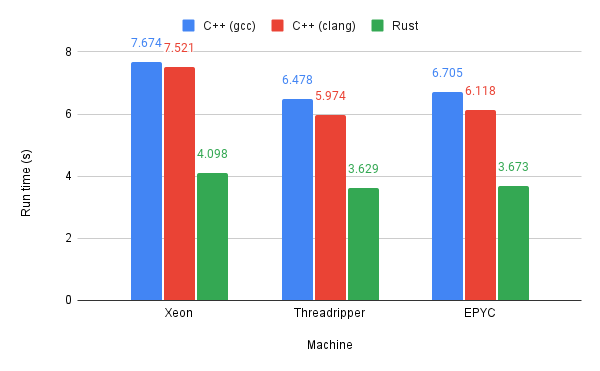}
    \caption{CBET sequential version runtime for 500 $\times$ 500 mesh}
    \label{fig:seq:500}
\end{figure}

\begin{figure}
    \centering
    \includegraphics[width=0.75\linewidth]{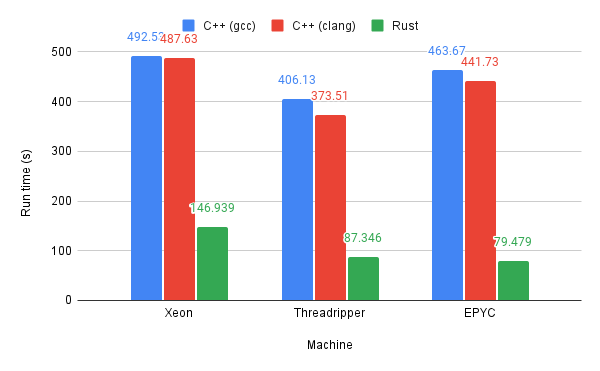}
    \caption{CBET sequential version runtime for 2500 $\times$ 2500 mesh}
    \label{fig:seq:2500}
\end{figure}

Figure~\ref{fig:seq:500} shows the performance results when simulating on a mesh with 500 zones in each dimension, and Figure~\ref{fig:seq:2500} shows the performance results when simulating on a mesh with 2500 zones in each dimension. For reference, the runtime of the MATLAB implementation with a 500-zone mesh, on the Xeon machine, is 36 seconds.

These results show that the Clang compiler produces an output that is up to 8.75\% faster than GCC. They also show that the Rust implementation is at least 1.6$\times$ as fast as the C++ version, in smaller cases, with performance gains of up to 5.6$\times$, in the case of the large mesh on the EPYC machine. Thus, the high level abstractions discussed in Sec~\ref{sec:impl:rust} come at a low enough performance cost to be viable in high-performance computing applications.


\subsection{Parallel Rust}

\begin{figure}
    \centering
    \includegraphics[width=0.75\linewidth]{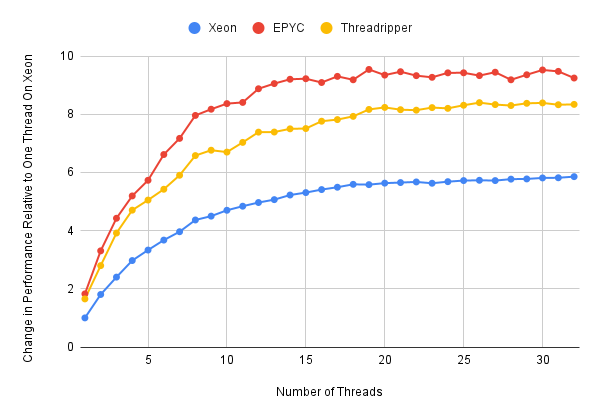}
    \caption{Multi-Threaded Performance on a 2500x2500 Mesh}
    \label{fig:strong-scaling-2500}
\end{figure}

Table~\ref{tab:parallel} shows the execution time of the parallel version on the same machines used above, with the same problem that the sequential versions of the simulator are tested. By default, Rayon uses the number of logical threads that are provided, so for Xeon, Threadripper, and EPYC, the number of threads are 36, 64, and 32 respectively. In order to test the strong scaling performance of the various machines available, we also tested the problem, for both mesh sizes, at different numbers of threads in the thread pool (capped at 32 threads). Figure~\ref{fig:strong-scaling-2500} shows the strong scaling performance of the simulator relative to using a single thread on the Xeon machine.

\begin{table}[h!]
\caption{Rust parallel version runtime}
\begin{center}
    \begin{tabular}{ |c|c|c| }
        \hline
        machine & 500x500 & 2500x2500 \\
        \hline
        Xeon & 1.344s & 24.926s \\
        \hline
        Threadripper & 1.313s & 17.727s \\
        \hline
        EPYC & 1.347s & 15.600s \\
        \hline
    \end{tabular}
\label{tab:parallel}
\end{center}
\end{table}

\section{Performance Profiling and Analysis}
\label{sec:perf:prof}
\subsection{Profiling}

In order to investigate the causes of the performance increase in the Rust implementation, we use the Linux perf tool to record performance information about the two implementations. The perf tool was run on the Xeon machine, testing each version with a mesh size of 500x500. For the C++ version, the Clang compiler was used, due to the higher performance of its output. Table~\ref{tab:perf-stats} describes the number of events recorded for the execution of the program. It shows that while the cache miss ratio for the Rust program is worse than the C++ program, the C++ program has many more cache accesses overall than the Rust program. Furthermore, the C++ program has many more branches than the Rust program.

To make the entire picture complete, we also profile the MATLAB program execution.
According to the profiling result, the C++ version improves the overall execution time by 4.8$\times$ compared with the MATLAB version.
The speedup can be explained by two differences.
First is the overhead associated with using an interpreted language.
To solve the same problem, MATLAB program spends 6.6$\times$ more cycles and issues 5.7$\times$ more memory accesses.
Second is the explicit memory management that enables better memory reuse.
The overall L1D cache misses ratio is reduced from 4.6\% to 3.9\%.

The perf tool was also used to record samples of the call graph, in order to determine which steps of the program are more performance intensive. Tables~\ref{tab:perf-cycles}-\ref{tab:perf-branch} describe the percentage of the samples recorded while executing each of the main steps of the CBET program, with samples recorded at a constant frequency, as well as samples recording cache loads and branches. These data show that the ray tracing step specifically is much more computationally expensive in the C++ implementation than the Rust implementation.

\begin{table}
    \caption{Occurrence of selected events, as measured by the perf tool, in billion events.}
    \begin{center}
    \begin{tabular}{ |c|c|c|c|c|c| }
        \hline
        event & MATLAB & C++ (clang) & Rust & C++/Rust \\
        \hline
        cycles & 215.931 & 32.624 & 18.229 & 1.79$\times$ \\
        L1-dcache-loads & 98.922 & 17.235 & 6.420 & 2.69$\times$ \\
        L1-dcache-load-misses & 4.527& 0.6796 & 0.4432 & 1.53$\times$ \\
        L1-dcache-stores & 58.954 & 10.103 & 2.444 & 4.13$\times$ \\
        branches & 65.708 &15.475 & 3.663 & 4.23$\times$ \\
        branch-misses & 0.534 & 0.0498 & 0.0206 & 2.42$\times$ \\
        \hline
    \end{tabular}
    \end{center}
\label{tab:perf-stats}
\end{table}

\begin{table}
    \caption{Occurrence of selected events, as measured by the perf tool, in billion events.}
    \begin{center}
    \begin{tabular}{ |c|c|c|c|c|c|c| }
        \hline
        event & MATLAB & C++ & Rust & MATLAB/C++ & C++/Rust \\
        \hline
        cycles & 215.931 & 32.624 & 18.229 & 6.62$\times$ &1.79$\times$ \\
        L1d loads & 98.922 & 17.235 & 6.420 & 5.74$\times$& 2.69$\times$ \\
        L1d load misses & 4.527& 0.6796 & 0.4432 & 6.67$\times$& 1.53$\times$ \\
        L1d stores & 58.954 & 10.103 & 2.444 & 5.84$\times$& 4.13$\times$ \\
        branches & 65.708 &15.475 & 3.663 & 4.25$\times$& 4.23$\times$ \\
        branch-misses & 0.534 & 0.0498 & 0.0206 & 10.72$\times$ & 2.42$\times$ \\
        \hline
    \end{tabular}
    \end{center}
\label{tab:perf-stats1}
\end{table}

\begin{table}
    \caption{Percentage of samples taken by the perf tool at 10,000 Hz in each section of the CBET code.}
    \begin{center}
    \begin{tabular}{ |c|c|c|c| }
        \hline
        section & MATLAB & C++ & Rust \\
        \hline
        Ray tracing - child ray & \multirow{2}{*}{24.10\%} & 26.75\% & 6.83\% \\
        Ray tracing - parent ray & & 25.42\% & 8.92\% \\
        CBET calculation & 65.61\% & 41.92\% & 78.30\% \\
        \hline
    \end{tabular}
    \end{center}
\label{tab:perf-cycles}
\end{table}

\begin{table}
    \caption{Percentage of samples taken by the perf tool for every 100,000 ``L1-dcache-loads'' events}
    \begin{center}
    \begin{tabular}{ |c|c|c| }
        \hline
        section & C++ & Rust \\
        \hline
        Ray tracing - child ray & 28.90\% & 11.65\% \\
        Ray tracing - parent ray & 38.26\% & 11.98\% \\
        CBET calculation & 31.79\% & 74.20\% \\
        \hline
    \end{tabular}
    \end{center}
\label{tab:perf-cache}
\end{table}

\begin{table}
    \caption{Percentage of samples taken by the perf tool for every 100,000 ``branches'' events}
    \begin{center}
    \begin{tabular}{ |c|c|c| }
        \hline
        section & C++ & Rust \\
        \hline
        Ray tracing - child ray & 37.01\% & 19.46\% \\
        Ray tracing - parent ray & 43.90\% & 20.61\% \\
        CBET calculation & 18.00\% & 56.09\% \\
        \hline
    \end{tabular}
    \end{center}
\label{tab:perf-branch}
\end{table}

\subsection{Analysis}

One possible explanation for this discrepancy is the data layout. The C++ implementation stores the data associated with crossings between rays and meshes in multiple arrays, with each point of data associated with a particular crossing stored at the same index in a separate array. The Rust implementation stores all of the data associated with a crossing in a struct, with each ray having a separate vector of crossing structs. However, this difference does not explain the fact that launching a child ray is also more expensive in the C++ version, despite the fact that launching the child ray does not save crossing information. Furthermore, it does not explain the difference in the number of branches, which would not increase so dramatically due to a different data layout.

Differences may be explained by more aggressive compiler optimizations on the part of Rust. However, understanding how compiler optimizations affect the performance of CBET is outside of the scope of this paper, due to the complexity in the way that compilers apply optimizations. For example, a study of LLVM optimization~\cite{delaTorre+:2018} passes found that their effect on performance is small unless examined in tandem with each other, a process that involves over a million different compilations. Furthermore, while LLVM backs both Rust and Clang, both compilers may also apply optimizations at other steps in the compilation process.

\section{Related Work}
\label{sec:related}

There are a few works that explore the benefit of using Rust for HPC applications.
Costanzo et al.~\cite{Costanzo+:CLEI21} study a scientific application named N-body, and find that Rust greatly reduces programming effort while providing similar performance compared to C.
They find that Rust is easier to be parallelized, has better readability, and needs fewer lines of code to express the same functionality.
Rotter et al.~\cite{Rotter+:CSCI22} further analyze a special implementation of the N-body problem, and find that while the C/C++ version takes less time to finish for small problems, Rust potentially has better scalability at larger problem scales. This finding is consistent with our findings, as the performance improvement of Rust over C++ is greater in the larger problem case.

\section{Conclusion}
\label{sec:conc}

We implement a simulation of the Cross Beam Energy Transfer interaction in C++ and in Rust, and compare their performance. Then, we measure the performance of a parallelized version of the Rust implementation that takes advantage of CPU multi-threading. We show that Rust can offer similar or even better performance for this task. Further, we show that parallel Rust code can be written easily and safely, with additional performance gains. Thus, we argue that Rust is a strong contender for high performance scientific computing applications.

Future work for Rust in scientific computing could involve further comparisons between code written in Rust and in other programming languages, both for performance and for ease-of-use.
However, Rust does not come without its limitations. For massively parallel workloads that may want to take advantage of GPUs, which can run thousands of threads at once, Rust may fall short, as GPU API bindings for Rust, such as CUDA, are still in their early development stages.
Furthermore, for Rust to be considered a more viable programming language in the scientific computing world, more libraries used in scientific computing should be written for Rust, or have Rust compatibility layers (as Rust has foreign function interfaces for C and C++).

There is more work to be done for CBET simulations as well, with the eventual goal of being able to simulate three-dimensional laser systems of dozens of lasers as efficiently as possible. The simulator computes intensities at one moment in time, so it must be able to run thousands of times with different mesh properties for each step of a full simulation of a laser system. In the future, optimizations for current implementations, as well as massively parallel implementations that take advantage of GPUs and multiple CPUs will be explored.

\section*{Acknowledgement}

The authors wish to thank Dylan McKellips, Jack Cashman, Giordan Escolona, and Woody (Yanhui) Wu for proof reading the paper and the anonymous reviewers of CASCON 2024 for the helpful comments and suggestions. The work is supported in part by the National Science Foundation (Contract No. SHF-2217395, CCF-2114319). This material is based upon work supported by the US DOE OFES under Award No. DE-SC0017951, US DOE OFES INFUSE program under Award No. DE-SC0024460, and US DOE NNSA University of Rochester “National Inertial Confinement Fusion Program” under Award No. DE-NA0004144.

\bibliographystyle{ieeetr}
\bibliography{refs}

\end{document}